Electron-Lattice Systems in Weak Gravitation: The Schiff–Dessler Problem

Timir Datta, and Michael Wescott

Physics and Astronomy department

University of South Carolina

Columbia, SC 29208

And

Ming Yin

Physics/Engineering Department

Benedict College

Columbia, SC 29204

**Overview:**

The behavior of composite matter in external fields can be very reveling. The quantum mechanical problem of a material object (test mass) placed in a uniform (weak) gravitational field, $g$, was considered by many authors starting with Schiff [Phys. Rev. **151**, 1067 (1966)]. Depending on the theoretical treatment opposing results of gravity induced (electric) field $E_g$ have been reported. In the Schiff model [L.I. Schiff, PRB, **1**, 4649 (1970)] $E_g$ is predicted to be oriented anti-parallel (with reference to $g$). On the other hand it is found to be parallel in the elastic lattice model [A. J. Dessler et al, Phys.Rev, **168**, 737, (1968); Edward Teller, PNAS, **74**, 2664 (1977)]. Surprisingly, this contradiction has been largely overlooked by modern researchers. Here an experimental test is suggested. We also reason that advanced density functional type calculations can provide valuable guidance.

**Introduction:**

It is well known that in many situations the interaction that is weaker in strength may determine bulk qualitative response of a system. For example in a system with free charges electromagnetic forces are far stronger than gravity. So gravitation may be "negligibly small" but still plays a critical role[1] in the equilibrium of gaseous plasma in the upper atmosphere of the earth or in stars[2]. Such behavior is not restricted to gasses, even in a condensed system such as a conducting solid the weaker magnetic force produces Hall Effect. Discovered by Edwin Hall (1879) this effect led to the conclusion that contrary to Maxwell the mobile charge carriers as opposed to the body of the solid itself, are the recipient of the (Lorentz) force.

Influence of gravity on conductors is interesting. This is a quantum many-body system where the negatively charged electrons behave as nearly-free fluid and the rest of the charges, the positively charged cores are arranged in a lattice. The problem is to find the equilibrium state of the conduction electrons plus the atomic cores under gravity. Schiff was amongst the first to tackle this problem.

**Effect of Gravity:**

In the Schiff model[3-5] the Hamiltonian of the conductor in the gravity field $g$, is considered to be:

$$H = H_0 + H_g + V \qquad \ldots 1$$

where, the Hamiltonian $H_0$ is that of the solid in absence of gravity and $V$ accounts for the supporting constraints. $H_g$ is the gravitational potential energy measured from the x-y plane, i.e.,

$$H_g = g[m_e \sum_i z_i + M_c \sum_j z_j] \qquad \ldots 2$$

Here, g is the free fall acceleration, $m_e$ the electron mass and $M_c$ is the core mass. The electric field operator $E(R)$ is given by,

$$E(R) = -\sum_i \frac{q(R-r_i)}{R^3_i} + \sum_j \frac{Zq(R-r_j)}{R_j^3} \qquad \ldots 3$$

The gravitational force causes the electrons to sink towards to the bottom. As in Hall Effect when sufficient charge segregation has taken place an internal electric field, $E_g$, builds up. Reasoning that the massive ion cores in the metal would not participate in the dynamical behavior of the solid expectation value of $E(R)$ was shown to be the free electron value of $-(m_e/q)g$, q is the electronic charge. From here on we will call $E(R)$, the gravitationally induced field $E_g$.

This model essentially neglects the gravitational compression of the core lattice. For classification purposes and to help differentiae amongst the various treatments henceforth we will call this a rigid lattice model. Because at equilibrium the down ward force (due to gravity) on the negatively charged electron has to be balanced by an upward "gravity induced" electric force, the field $E_g$ must point vertically down.

The essential predictions of the rigid lattice model can be summarized as: under gravity the electrons in the system redistribute to reach equilibrium (i) and attain zero acceleration, conversely (ii) a positron must fall with acceleration of 2g. Schiff's model describes quantum sedimentation of the conduction electrons in a perfectly rigid positive background. Note, although not considered by these authors, a consequence of charge segregation is that the conductor would acquire an electric dipole moment $P_g$, which in this model points vertically up.

Shortly after the publication of Schiff's first paper, a controversy started over the incorrect assessment of the lattice contribution in the Schiff model. Calculations by Dessler etal[5] and others included better accounting of lattice compressibility effect. In the Dessler treatment electrochemical potential of the electrons was held constant and the inhomogeneous (vertical) charge density was maintained by $E_g$. These calculations predicted that the induced electric field is in the opposite direction of gravity that is up ward, and is much bigger. Indeed, with the inclusion of the core contributions $E_g \sim g/q(M_c/m_e)$. As the authors point out that the value of $E_g$, even close to the surface[6,7] will be daunting because unlike µ, the Chemical potential of the metal both work function and field are strongly influenced by local crystallographic and surface consideration[8].

There have been several other discussions of this problem [9-11]. In general for elastic models the estimate for the induced field come out to be the same order of magnitude (~$10^5$) higher than that in the rigid system. Furthermore, in the compressible lattice model the electrons actually rise against gravity and the whole object acquires dipole moment $\boldsymbol{P}_g$, in the direction of gravity.

Physically, in an elastic system, the gravitational acceleration due to the earth produces a far greater compression of the massive lattice. In the bottom of the conductor this compression creates a bigger positive charge density (background), than that of the far lighter and less compressible (Fermionic) free-electrons. Also, the conductor as whole gets polarized with the positive pole at the bottom.

Teller considered insulating dielectric matter with two different ionic masses, $M_+$ and $M_-$ respectively[12]. From the calculations of electric dipole moment of a rapidly rotating dielectric Teller predicted the generation of magnetic fields near the object. Even for systems with the largest ionic mass ratio ($M_+/M_-$) it is not possible to reach the high value of ($M_c/m_e$) so very rapid rotation will be required to create large enough acceleration to produce detectable signals. He argued that such acceleration measurements can be important in the investigations of ferroelectric and related phenomena. Unfortunately, surface field measurements of rotating objects are difficult especially at high angular velocities and the technique has received little attention but research with rotating conductors is active[13-16].

We reason that it is important to study the gravitational response of conducting materials. It appears that not much theoretical effort has been directed to this problem. Perhaps calculations using the density functional theory (DFT) approach will provide quantitatively testable, materials property dependant predictions. Even with elastic enhancement, in engineering terms the induced field is rather small $\boldsymbol{E}_g \sim 1$ µV/m. There has been some experimental report[17] on gravitational effects on the emf of electrochemical cells, but we reason that it is better to measure $P_g$ rather than $E_g$, The idea is to take advantage of cumulative built up of the effect of gravity over the total sample and avoid complications due to surface effects.

**Experimental Test:**

We propose a simple experiment to directly measure the gravity induced dipole moment $P_g$ of a macroscopic (blue ball in figure 1) spherical test mass. $P_g$ can be determined from measurements of the total weight force ($F$) when the sample is subjected to a non-uniform electric field. Their respective forces as well as the time cycled bias voltage applied on the inhomogeneous capacitor are also shown in figure 1.

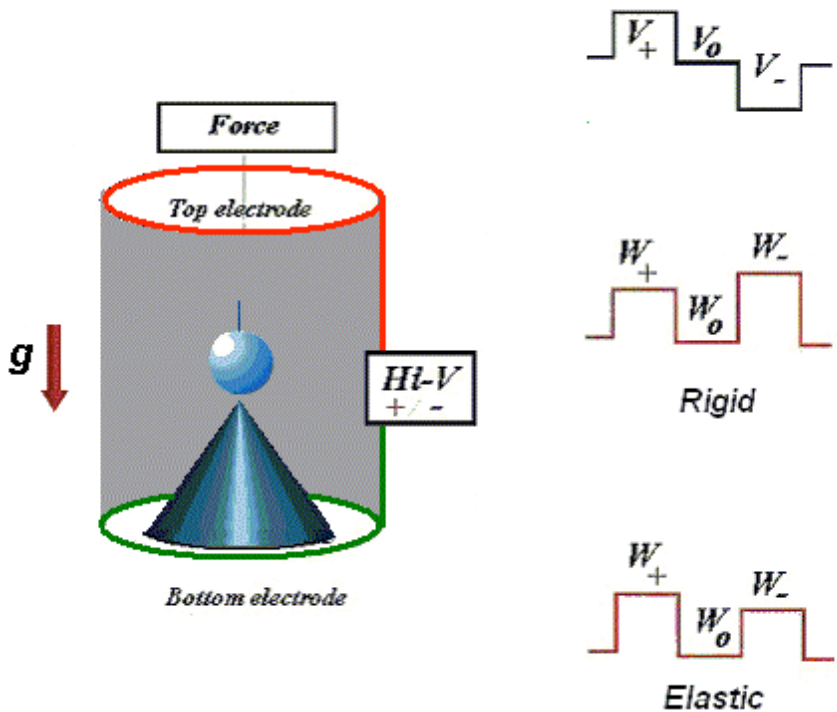

**Figure 1:** A schematic of the proposed experiment to determine the gravity induced electric dipole moments ($P_g$) of a test mass.

Let us express the total weight force $W_+$ ($W_-$) under positive (negative) bias as a sum of (i) the weight force $W_o$; (ii) the force due to the induced electrostatic dipole ($P_{elect}$) and (iii) that of the gravity induced dipole ($P_g$) as follows

$$W_+ = -W_o + [-|P_{elect}| - |P_{grav}|] \bullet \left|\frac{dE}{dz}\right| \qquad \ldots 4$$

and

$$W_- = -W_o + [-(-|P_{elect}|) - |P_{grav}|] \bullet \left(-\left|\frac{dE}{dz}\right|\right) \qquad \ldots 5$$

In the above equations, we have retained the following experimental constraints: irrespective of bias, the weight force is always down (negative); $P_{elect}$ flips over along with the bias but always remain parallel to the applied field; the gradient of the electric field is positive when the positive bias is applied to the top electrode; finally $P_g$ depends only on gravity and is independent of the applied external electric field. Also, $P_g$ remains fixed in space either parallel or anti-parallel to gravity (z-axis) depending on which of the two theoretical models is correct. If the experimental observation indicates a larger apparent weight with positive bias then $P_g$ points down and vice versa if the weight at positive bias is smaller.

From the above equations we can also separate the electrostatic ($F_{elec}$) and the gravitational ($F_{grav}$) force components as follows:

$$F_{elct} = \frac{1}{2}[W_+ + W_- - 2 \bullet W_0] = |P_{elct}| \bullet \left|\frac{dE}{Dz}\right| \qquad \ldots 6$$

and

$$F_{grav-elect} = \frac{1}{2}[W_+ - W_-] = |P_{grav}| \bullet \left|\frac{dE}{dz}\right| \qquad \ldots 7$$

From the knowledge of the experimental parameters such as sample volume, applied field gradient vector and the measured values of the weight forces under zero, top positive and top negative biases, $W_0$, $W_+$ and $W_-$ respectively, one can determine the direction and magnitude of $P_g$.

A direct measurement of the gravity-induced dipole moment $P_g$ will also fix $E_g$ and hence resolve the Schiff- Dessler controversy.

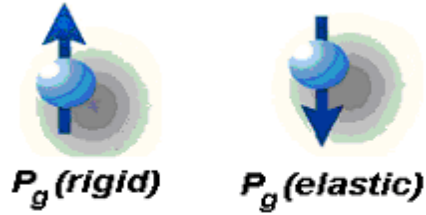

**Figure 2:** The direction of the gravitationally induced dipole moments in the rigid and elastic models.

As described above, the difference between Schiff's treatment and the more realistic models is clearly quantitative. In the later or elastic models, the effect of gravity is five orders of magnitude larger and entirely due to the massive core lattice. However the most striking difference is qualitative $E_g$ is directed upwards opposite to Shiff's prediction. If Schiff's original prediction is correct then the electrons will sink to the bottom and the corresponding electric dipole moment $P_g$, points vertically up. Like wise in the elastic case $P_g$, points vertically down (Figure 2).

**Future Modifications:**

It will also be interesting to study the effects of material properties, especially with theoretical guidance via DFT prediction for viz, elasticity and mass dependence by varying the density of the alloy. Gravity enters the Hamiltonian of problem through the $H_g$ part which as shown in equation 2 is linear in the mass terms $m_e$ and $M_c$. However especially in time dependent situations band structure effects are likely to be critical. Comparison between very heavy fermionic conductors vis-à-vis very light effective mass conductors will also be interesting. In other iterations, similar to the electron-positron free-fall in a Fairbank type experiment[18], the behaviors of n and p-type semiconductor test masses can be compared.

**Practical Implications:**

Results of the proposed experiment can impact gravity related experiments including the Schiff gyroscope. Significant electric dipole moment on metal particles can give rise to considerable electro static forces. For example, the controversy between Millikan and Errenhaft regarding the discrete nature of electronic charge can be accounted for by the differences between gravity and electric forces on Millikan's oil drops and metal dust used by Ehrenhaft[19,20]. The equilibrium charge distribution due to gravity on spherical (dielectric) compared with that of random shapes of (conducting) metal particles are not the same; also the same applies to the report of "free quark" on metal spheres[17]. As indicated earlier the dipole moment is proportional to and is along the direction of gravity, so it is possible to envision semiconductor based devices that can be useful in high precision gravity detection.

**Acknowledgments:**

A shortened version of this paper was submitted as an essay to the gravity foundation in 2008.